\documentclass[aps]{revtex4}
\usepackage{graphicx}

\newcommand{\tr}{{\rm tr}}

\newcommand{\dslash}{{\partial \hskip -6pt /}}

\begin{document}

\title{Nontopological Finite Temperature Induced Fermion Number}
\author{I. J. R. Aitchison$^{(1)}$ and G. V. Dunne$^{(1,2)}$}
\affiliation{$^{(1)}$ Department of Theoretical Physics, Oxford
University, 1 Keble Rd., Oxford OX1 3NP, U.K.\\
$^{(2)}$ Research Centre for the Subatomic Structure of Matter,
University of Adelaide, S.A. 5005, Australia}


\vskip 1cm

\begin{abstract}
We show that while the zero temperature induced fermion number in a chiral
sigma model background depends only on the asymptotic values of the chiral 
field, at finite temperature the induced fermion number depends also on the 
detailed shape of the
chiral background. We resum the leading low temperature terms
to all orders in the derivative expansion, producing a simple result  that can be
interpreted physically as the different effect of the chiral background on virtual
pairs of the Dirac sea and on the real  particles of the thermal plasma.  By
contrast, for a kink background,  not of sigma model form,  the finite T induced
fermion number is temperature dependent but topological.

\end{abstract}

\maketitle



The phenomenon of induced fermion number due to the interaction of fermions with
topological backgrounds (e.g., solitons, vortices, monopoles, skyrmions) has many
applications ranging from polymer physics to particle physics
\cite{jr,ssh,gw,wilczek,gj,eric,diakonov,niemi}. The original fractional fermion
number result of Jackiw-Rebbi \cite{jr} has a deep connection with the existence of
spinless charged excitations in polymers \cite{ssh}. The adiabatic analysis of
Goldstone-Wilczek \cite{gw} in systems without conjugation symmetry has important implications for bag
models \cite{gj}, monopoles, and sigma models, which provide effective field theory
descriptions of systems ranging from condensed matter, to AMO, to particle and
nuclear physics \cite{weinberg}. The induced fermion  number is related to the
spectral asymmetry of the relevant Dirac operator,  and mathematical results
concerning index theorems \cite{niemi}  relate the fermion number to asymptotic
topological properties of the background. At finite  temperature, the situation is
less clear. In several examples
\cite{ns,babu,cp,goldhaber,monopole}, the fermion number is known to be  temperature
dependent, but is still topological in the sense  that the only dependence on the
background field is through its asymptotic properties. In this Letter, we present a
simple physical case for which this is not true : in a $1+1$ dim chiral sigma model,
the finite temperature induced fermion number depends on the detailed structure of
the background. This contradicts a previous analysis \cite{midorikawa} and claim
\cite{nnp} that the finite T fermion number is in general a topological  quantity.
We give a simple physical explanation of the origin of the  nontopological
dependence. Our analysis has been motivated in part by the results of
\cite{pisarski} concerning the T dependence of anomalous amplitudes in nuclear
decays.

Consider an abelian model in $1+1$ dimensions with fermions interacting via
scalar and pseudoscalar couplings to two bosonic fields $\phi_1$ and $\phi_2$.
For the purposes of this paper $\phi_1$ and $\phi_2$ can be considered as
classical external fields. The Lagrangian is
\begin{equation}
{\cal L}=i\,\bar{\psi}\dslash\,\psi -\bar{\psi}\left(\phi_1+i\,
 \gamma_5\,\phi_2\right)\psi
\label{lag}
\end{equation}
There are two especially interesting physical cases:

(i) kink case \cite{jr} :
\begin{equation}
\phi_1=m\qquad\qquad {\rm and} \qquad \qquad \phi_2(\pm\infty)=\pm\hat{\phi_2}
\label{kink}
\end{equation}

(ii) sigma model case \cite{gw} : 
\begin{equation}
\phi_1^2+\phi_2^2=m^2
\label{chiral}
\end{equation}
In the sigma model case (\ref{chiral}), the interaction term in the 
Lagrangian (\ref{lag}) can be expressed as
\begin{equation}
m \bar{\psi}\, e^{i\gamma_5 \theta}\, \psi=m \bar{\psi}\left(
\cos\theta+i\gamma_5 \sin\theta\right)\psi
\label{exp}
\end{equation}

\begin{figure}[b]
\includegraphics[scale=0.5,angle=270]{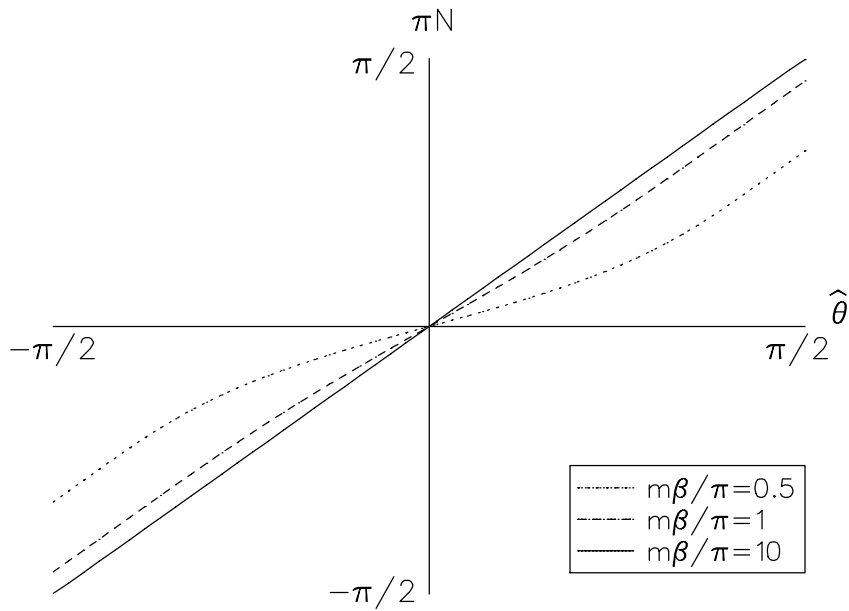}
\caption{Plots of $\pi N$, where $N$ is the finite temperature  fermion number
(\protect{\ref{kinkq}}) for the kink case  (\protect{\ref{kink}}), as a function of
$\hat{\theta}$. These plots are for 
$m\beta/\pi$ taking values $0.5$, $1$, and $10$, as labelled. As  $T\to 0$,  note
that $\pi N\to\hat{\theta}$, as in (\protect{\ref{gwcharge}}).}
\label{f1}
\end{figure}

At $T=0$, both these cases have  an 
induced topological current $J^\mu\equiv <\bar{\psi}\gamma^\mu\psi>$ given by 
\cite{gw}
\begin{equation}
J^\mu=\frac{1}{2\pi}\,\epsilon^{\mu\nu}\partial_\nu \theta +\dots
\label{gwc}
\end{equation}
where the angular field $\theta$ is defined by 
$\theta\equiv {\rm arctan}(\phi_2/\phi_1)$.
The dots in (\ref{gwc}) refer to higher derivative terms, which are
all of the form of a total derivative of $\theta$ and its derivatives 
\cite{wilczek}. Thus, in particular, the induced fermion number, 
$N\equiv\int dx\, J^0$, is
\begin{equation}
N=\frac{1}{2\pi}\int_{-\infty}^\infty dx\, \theta^\prime
=\frac{1}{\pi}\,\hat{\theta}
\label{gwcharge}
\end{equation}
where $\pm\hat{\theta}$ are the asymptotic values of $\theta(x)$ at 
$x=\pm\infty$. The fermion number $N$ is topological as it depends only
on $\hat{\theta}$, not on the detailed shape of $\theta(x)$. The conjugation
symmetric case of Jackiw and Rebbi \cite{jr} is obtained by taking $m\to 0$
in the kink case (\ref{kink}), in which case $N\to \pm\frac{1}{2}$.

At nonzero temperature, the induced fermion number for a static background is 
\cite{niemi,monopole}
\begin{equation}
N=-\frac{1}{2}\int_{\cal C}\,\frac{dz}{2\pi i}\, \tr\left(\frac{1}{H-z}\right)
{\rm tanh}\left(\frac{\beta z}{2}\right)
\label{tcharge}
\end{equation}
where $\beta=1/T$ is the inverse temperature, and $\tr (\frac{1}{H-z})$ is 
the resolvent of the Dirac Hamiltonian $H$. The contour ${\cal C}$ is 
$(-\infty+i\epsilon,+\infty+i\epsilon)$ and 
$(+\infty-i\epsilon,-\infty-i\epsilon)$.
By considering static backgrounds we
avoid the well-known complications of finite temperature calculations
in non-static backgrounds \cite{das}. The technical part of the
calculation of the induced fermion number (\ref{tcharge}) is the computation
of the resolvent of $H$. Once this is done, the induced fermion number may
be expressed as an integral representation, or as a sum
by deforming the contour in (\ref{tcharge}) around the simple poles of the 
tanh function. For static backgrounds $\phi_1(x)$ and $\phi_2(x)$ in 
(\ref{lag}), the Dirac Hamiltonian is
\begin{equation}
H=-i\gamma^0\gamma^1\nabla +\gamma^0 \phi_1(x)+i\gamma^0\gamma_5 \phi_2(x)
\label{ham}
\end{equation}
where $\nabla\equiv\frac{d}{dx}$, and we will work with the Dirac matrices
$\gamma^0=\sigma_3$, $\gamma^1=i\sigma_2$, and $\gamma^5=-\sigma_1$.
Also, note that only the {\it even} part (in terms of the argument $z$) 
of the resolvent $\tr (\frac{1}{H-z})$ contributes to the induced fermion
number $N$ in (\ref{tcharge}). (This is most easily seen by deforming the 
contour around the poles of the tanh function.)


Consider first the kink case in (\ref{kink}). Then the even part of the 
resolvent can be computed exactly using a trace identity which is a special
case of the Callias index theorem \cite{callias,niemi,ns} (alternatively, it 
can be derived in a more elementary manner as an exact 
resummation of a SUSY derivative expansion \cite{dunne}) :
\begin{eqnarray}
\left[\tr\left(\frac{1}{H-z}\right)\right]_{\rm even} 
&=& \tr\left({m\over -(\nabla+\phi_2)(\nabla- \phi_2)+m^2-z^2}\right)- 
\tr\left({m\over - (\nabla-\phi_2)(\nabla+\phi_2)+m^2- z^2}\right)\nonumber\\
&=& {- m \hat{\phi}_2 \over (m^2-z^2)\,\sqrt{m^2+\hat{\phi}_2^2-z^2}}
\label{trace}
\end{eqnarray}
Then the induced fermion number (\ref{tcharge}) for the kink case (\ref{kink})
is
\begin{equation}
N= \frac{2}{\pi}\, \left(\frac{m\beta}{\pi}\right)^2 {\rm sin}\hat{\theta}\,
\sum_{n=0}^\infty {1\over ((2n+1)^2 +(\frac{m\beta}{\pi})^2) \sqrt{(2n+1)^2
\cos^2\hat{\theta}+(\frac{m\beta}{\pi})^2}}
\label{kinkq}
\end{equation}
where $\hat{\theta}\equiv{\rm arctan}(\hat{\phi_2}/m)$. This result is 
consistent with previous analyses \cite{ns}, although these were
much less explicit. The induced fermion number (\ref{kinkq}) is plotted 
in Fig. 1 as a function of $\hat{\theta}$ for 
various values of temperature. As $T\to 0$, this result reduces 
smoothly to the zero temperature result (\ref{gwcharge}). Despite its
complicated form, the nonzero 
temperature result (\ref{kinkq}) is still topological as it only 
refers to the background through $\hat{\theta}$.


In the sigma model case (\ref{chiral}), the trace identity formulae 
(\ref{trace}) 
do not apply. Another approach is needed to evaluate the resolvent. One such 
approach is the derivative expansion \cite{ian},
in which we assume that the spatial derivatives of the background fields are
small compared to the fermion mass scale $m$. In other words, the backgrounds
$\phi_1(x)$ and $\phi_2(x)$  are assumed to be
slowly varying on the scale of the fermion Compton wavelength.
Returning to the general Hamiltonian (\ref{ham}), the derivative 
expansion can be obtained by separating $H^2$ as
\begin{equation}
H^2=\left(\matrix{-\nabla^2+\phi_1^2+\phi_2^2&0\cr 0&  
-\nabla^2+\phi_1^2+\phi_2^2}\right)+\left(\matrix{\phi_2^\prime& 
i \phi_1^\prime\cr -i \phi_1^\prime& -\phi_2^\prime}\right)
\label{deriv}
\end{equation}
and then expanding $\tr\left(\frac{1}{H-z}\right)=
\tr\left((H+z)\frac{1}{H^2-z^2}\right)$ in powers of derivatives. A simple 
calculation to first order yields:
\begin{equation}
\left[ \tr\left(\frac{1}{H-z}\right)\right]_{\rm even}=-\frac{1}{2}
\int_{-\infty}^\infty dx\, {(\phi_1\,\phi_2^\prime -\phi_2\, \phi_1^\prime)
\over (\phi_1^2+\phi_2^2-z^2)^{3/2}}+\dots
\label{first}
\end{equation}
where the dots refer to terms with three or more derivatives.

In the kink case (\ref{kink}), where $\phi_1=m$ is 
constant, this first order calculation actually reproduces the {\it exact} 
trace identity result (\ref{trace}).  
But in the sigma model case (\ref{chiral}), where $\phi_1^2+\phi_2^2=m^2$ is 
a constant, the first order derivative expansion result (\ref{first}) implies
that:
\begin{equation}
\left[ \tr\left(\frac{1}{H-z}\right)\right]_{\rm even}=-
\frac{m^2}{2(m^2-z^2)^{3/2}} \,\int_{-\infty}^\infty dx\, \theta^\prime +\dots
\label{chiralfirst}
\end{equation}
So, to first order in the derivative expansion, the induced fermion number
for the sigma model case is
\begin{equation}
N^{(1)}=\frac{1}{\pi}\left(\frac{m\beta}{\pi}\right)^2 \left(\sum_{n=0}^\infty 
\frac{1}{[(2n+1)^2+(\frac{m\beta}{\pi})^2]^{3/2}}\right)\, 
\int_{-\infty}^\infty dx\, 
\theta^\prime
\label{qfirst}
\end{equation}
which is simply the zero temperature answer (\ref{gwcharge}) multiplied by 
a smooth function of $T$. As $T\to 0$, this prefactor reduces to 
$\frac{1}{2\pi}$,
so the {\it full} zero temperature result (\ref{gwcharge}) is regained.
But at finite temperature, the first order (in the derivative expansion)
formula (\ref{qfirst}) for the sigma model case differs from the kink case
formula (\ref{kinkq}), even though 
each of (\ref{qfirst}) and (\ref{kinkq}) reduces to (\ref{gwcharge}) at $T=0$.

This raises the question of the higher order corrections to the derivative
expansion (\ref{first}). In the kink case (\ref{kink}), there are {\it no}
higher order corrections to the even part of the resolvent in (\ref{deriv}).
This is due to the special form of the Hamiltonian in the kink background,
which leads to the first order formula (\ref{first}) agreeing with the exact 
trace identity result (\ref{trace}). There can, of course, be higher
order corrections to the induced fermion number {\it density},
but these are all total (spatial) derivatives, and do not
contribute to the integrated induced fermion number, 
even at nonzero temperature.

\begin{figure}[b]
\includegraphics[scale=0.5,angle=270]{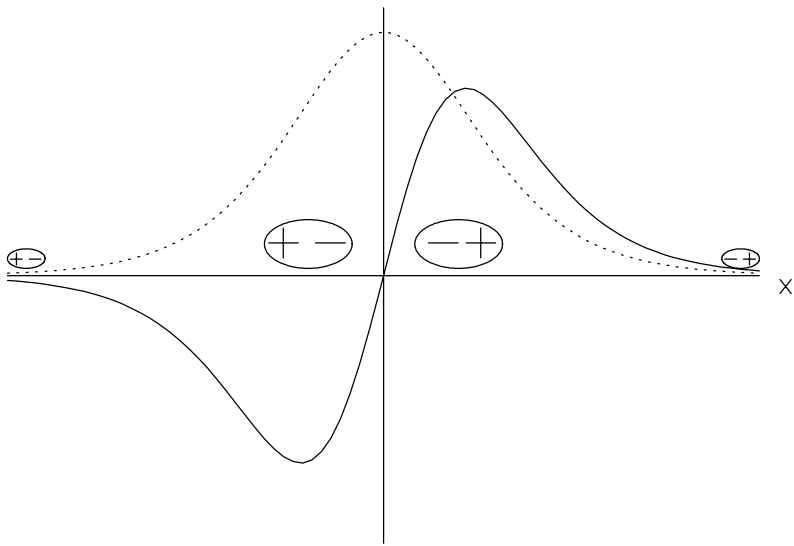}
\caption{For a kink-like chiral field $\theta(x)$ (dashed line), the 
corresponding electric field in (\protect{\ref{electric}}) has 
the form shown in the solid line, producing a vacuum polarization
charge  distribution localized near the kink center, roughly following
the dotted line $\theta^\prime$.}
\label{f2}
\end{figure}

But in the sigma model case (\ref{chiral}), where the trace identity does not 
apply, the situation is very different. Going to the next order in the 
derivative expansion, we find
\begin{eqnarray}
\left[ \tr\left(\frac{1}{H-z}\right)\right]_{\rm even}=
-\frac{m^2}{2(m^2-z^2)^{3/2}}\int dx\, \theta^\prime & -& 
\frac{m^2}{8(m^2-z^2)^{5/2}}
\int dx\, \theta^{\prime\prime\prime}\nonumber\\
&-&
\frac{m^2(4z^2+m^2)}{16 (m^2-z^2)^{7/2}} \int dx\, (\theta^\prime)^3 +\dots
\label{third}
\end{eqnarray}
where the dots refer to terms involving five or more derivatives.
For a chiral background with $\theta(x)$ approaching its asymptotic 
values exponentially fast, the term $\int dx\,\theta^{\prime\prime\prime}$ 
vanishes. But $\int dx\, (\theta^\prime)^3$ does not vanish. Thus, the first
order induced fermion number (\ref{qfirst}) acquires a third order correction:
\begin{equation}
N^{(3)}=\frac{\beta^2}{8\pi^3}\left(\frac{m\beta}{\pi}\right)^2\,
 \left(\sum_{n=0}^\infty {[-4(2n+1)^2+(\frac{m\beta}{\pi})^2] \over
[(2n+1)^2+(\frac{m\beta}{\pi})^2]^{7/2}}\right)\, \int dx\, (\theta^\prime)^3
\label{q3}
\end{equation}
This is {\bf not} just a function of the asymptotic value 
$\hat{\theta}$ of the chiral field $\theta(x)$; it also depends on the actual 
{\it shape} of $\theta(x)$. Thus, the induced fermion number is no longer
topological. This contradicts  \cite{midorikawa}, where it is stated that the 
first order derivative expansion contribution (\ref{qfirst}) is the full 
answer. However, the energy trace prefactor in (\ref{q3}) vanishes at $T=0$, 
so the 
nontopological third order contribution (\ref{q3}) vanishes at $T=0$. Thus, 
the nontopological nature of the finite temperature induced fermion number
is still consistent (at this order) with the topological nature of the zero 
temperature induced fermion number (\ref{gwcharge}).

We now turn to a physical explanation of why, in the sigma model case, the 
finite 
temperature induced charge is more sensitive to the background field than at 
zero temperature. Note first of all that the chiral background acts like a 
static but spatially
inhomogeneous electric field, as can be seen by making a local chiral
rotation \cite{wilczek}: 
$\psi\to\tilde{\psi}=e^{i\theta \gamma_5/2}\psi$. In
terms of these chirally rotated fields the Lagrangian (\ref{lag}), with 
interaction (\ref{exp}), becomes
\begin{equation}
{\cal L}=i\bar{\tilde{\psi}}\dslash \tilde{\psi}-
m \bar{\tilde{\psi}} \tilde{\psi}
-\bar{\tilde{\psi}}\, \gamma^0\, \frac{\theta^\prime}{2}\,\tilde{\psi}
\label{rotlag}
\end{equation}
(The chiral rotation leads to an anomalous Jacobian in the path 
integral, but this does not affect the induced fermion number.) Thus, 
the chiral field acts as an inhomogeneous 
$A_0(x)=\frac{1}{2}\theta^\prime(x)$, leading to an electric field
\begin{equation}
E(x)=\frac{1}{2}\,\theta^{\prime\prime}(x)
\label{electric}
\end{equation}
Given that $\theta(x)$ itself has a kink-like spatial profile, the electric 
field is such that it changes sign as a function of $x$, as shown in 
Fig. 2 (we choose $\theta^\prime>0$). This 
electric field acts on the Dirac sea to polarize the vacuum by aligning the 
virtual vacuum dipoles of the Dirac sea, producing a localized build-up of 
charge near the kink center. But at nonzero temperature, the 
electric field also has an effect on the thermal plasma, as we show below.



First, consider the full derivative expansion (\ref{first}) of the even part
of the resolvent, at low but nonzero 
temperature. At fifth order, there are three independent terms involving
$\theta^{\prime\prime\prime\prime\prime}$, $\theta^{\prime\prime\prime} 
(\theta^\prime)^2$, and $(\theta^\prime)^5$. The 
$\theta^{\prime\prime\prime\prime\prime}$ term vanishes when integrated
over $x$, but the other two terms are generally nonzero. However, as $T\to 0$
the $(\theta^\prime)^5$ term dominates the $\theta^{\prime\prime\prime} 
(\theta^\prime)^2$ term. Indeed, for low temperature, the dominant term 
with $(2l-1)$ derivatives in the derivative
expansion (\ref{first}) involves $(\theta^\prime)^{2l-1}$. 
Then, using the chirally rotated form (\ref{rotlag}) of the Lagrangian,
the dominant term at $(2l-1)^{\rm th}$ order is simply:
\begin{equation}
N^{(2l-1)}_{\rm dom}=\left(T \sum_{n=-\infty}^\infty\, \int \frac{dk}{2\pi}
{\tr([\gamma^0( p\hskip -5pt / +m)]^{2l})\over (p^2+m^2)^{2l}}\right)
\int dx\, \left(\frac{\theta^\prime}{2}\right)^{2l-1}
\label{lth}
\end{equation}
with Euclidean $p=(\omega_n,k)$ and $\omega_n=(2n+1)\pi T$ the Matsubara modes.

At zero temperature, all these terms $N^{(2l-1)}$ vanish, except for $l=1$. 
This fact is not obvious; it
involves highly nontrivial cancellations between terms in the expansion of 
the trace. But at nonzero temperature, all the terms in (\ref{lth}) are
non-vanishing. Moreover, they have a remarkably simple low temperature 
$(T\ll m)$ limit:
\begin{equation}
N^{(2l-1)}=\delta_{l,1}\,\int dx\,\frac{\theta^\prime}{2\pi}-
\sqrt{\frac{2mT}{\pi}}\, e^{-m/T}\, \frac{1}{(2l-1)!}\,
 \int dx\,  \left(\frac{\theta^\prime}{2T}\right)^{2l-1}+\dots 
\label{asymptotic}
\end{equation}
Thus, in the low temperature limit, we can resum the {\it entire} derivative 
expansion, to obtain the induced
fermion number in the sigma model case (\ref{chiral},\ref{exp}) :
\begin{equation}
N=\int dx\, \frac{\theta^\prime}{2\pi} -
\sqrt{\frac{2mT}{\pi}}\,\int dx\, e^{-m/T}\, 
{\rm sinh}\left(\frac{\theta^\prime}{2T}\right)+\dots
\label{resum}
\end{equation}
where the dots refer to subleading terms for $T\ll m$.

Several features of this result (\ref{resum}) deserve comment. First, note 
that at zero temperature, only the first term survives, producing the 
familiar result (\ref{gwcharge}) 
that the induced fermion number depends on the chiral field $\theta(x)$
only through its asymptotic value $\hat{\theta}\equiv\theta(\infty)$. 
At zero temperature, one can invoke Lorentz invariance to constrain the 
form of higher order corrections to (\ref{gwc}) to be total derivatives 
\cite{wilczek}, but these arguments do not apply at finite temperature. We
see this in (\ref{resum}): the temperature dependent corrections are not total 
derivatives of terms made from $\theta$ and its derivatives.
At nonzero temperature this shows clearly that the induced 
fermion number is nontopological - it depends also on the detailed shape 
of $\theta(x)$. Second, the resummed exponential factors 
$e^{-(m\mp \theta^\prime/2)/T}$ in (\ref{resum}) are 
consistent with the derivative expansion assumption that $\theta^\prime \ll m$.
Finally, the form of these exponential factors suggests an interpretation of 
the derivative expansion as an adiabatic change of the local Fermi level with
a local chemical potential $\mu=-\theta^\prime/2$, which
once again is only sensible in the derivative expansion regime where 
$\theta^\prime\ll m$.

To make this physical picture more precise, we can interpret the result 
(\ref{resum}) as follows. The first, topological, term refers to the induced 
charge coming from the polarization of the Dirac sea. This is temperature 
independent as the short-lived virtual ``electron-positron dipoles'' of the 
Dirac sea do not come to thermal equilibrium. The next term in (\ref{resum}) 
corresponds to the
induced charge arising from the response of the real charges in the thermal
plasma to the spatially inhomogeneous electric field (\ref{electric}). Indeed,
the linear response \cite{lebellac} of the plasma at low
temperature to such an electric field yields an induced fermion number density
\begin{equation}
\rho(x)=\int\frac{dk}{2\pi}\, f(x,k)
\label{ind}
\end{equation}
where the static distribution function $f(x,k)$ satisfies the Boltzmann
equation
\begin{equation}
v\, \frac{\partial}{\partial x}\, f(x,k)=- E(x) \frac{\partial}{\partial k}\, 
f(x,k)
\label{boltzmann}
\end{equation}
where $v=k/\sqrt{k^2+m^2}$. Regarding $\mu=-\frac{1}{2}\theta^\prime(x)$ as
a local chemical potential, (\ref{boltzmann}) is satisfied by local
Fermi particle and antiparticle distribution functions 
\begin{equation}
f_\pm(x,k)={1\over e^{\beta (\sqrt{k^2+m^2}\mp \mu)}+1}
\label{fermi}
\end{equation}
Inserting $f=f_+-f_-$ into (\ref{ind}), we obtain
precisely the second, nontopological, term in (\ref{resum}) in the low 
temperature limit.

At T=0, the fermion number may be defined as a sharp observable \cite{gk}; but at
$T>0$, thermal fluctuations introduce an  rms deviation. Thus,  the finite T fermion
number in (\ref{tcharge},\ref{resum}) is a thermal expectation value $\langle
N\rangle$, as in the monopole cases
\cite{cp,goldhaber,monopole}. We have estimated  $\langle N^2\rangle-\langle
N\rangle^2$, in the derivative expansion regime, in an analogous manner to the
computation presented here for $\langle N\rangle$. We find that the rms deviation
vanishes at T=0, but at nonzero T can be significant compared to the thermal shift
in (\ref{resum}). Details of this will be reported elsewhere.

To conclude, we comment briefly on possible implications of these results
for models in other dimensions for which there is an induced fermion number due
to some nontrivial background. In $2+1$ dimensions, fermions in a static 
magnetic background acquire an induced charge that is topological, 
expressed in terms of the net magnetic flux of the background. At 
finite temperature, the induced charge remains topological, but is multiplied 
by a smooth function of the temperature \cite{babu}.  
In $3+1$ dimensions, fermions in a static Dirac monopole background acquire 
an induced charge that is temperature dependent at finite $T$, but still
only depends on the background through the total magnetic charge and the 
self-adjoint extension parameter \cite{cp,monopole}. A more
interesting case is a static 't Hooft-Polyakov monopole background, which
has a characteristic size scale. Consider, for example, the coupling
\begin{equation}
{\cal L}_{\rm int}=\bar{\psi}\left( A\hskip -6pt /+\phi+i\gamma_5 m\right)\psi
\label{mon}
\end{equation}
where $\psi$ is an isodoublet fermion, $A_\mu$ is a static $SU(2)$ 
't Hooft-Polyakov monopole, and $\phi$ is the corresponding static
Higgs field. We have computed the finite temperature induced 
fermion number, using the $3+1$ trace identity used in the
zero temperature case \cite{manu}, and we find {\it precisely} the same
expression (\ref{kinkq}) as in the $1+1$ kink case, with the identification
$\hat{\theta}={\rm arctan}(\hat{\phi}/m)$, where $\hat{\phi}$ is the 
asymptotic value of the magnitude $|\phi|=\sqrt{\phi^a \phi^a}$ of the 
Higgs field. Given that (\ref{kinkq}) reduces to (\ref{gwcharge})
at $T=0$, this monopole result is consistent with the familiar zero
temperature result \cite{gw,manu} that the induced fermion number is 
proportional to $\hat{\theta}$ \cite{comment}. 
The $3+1$ dimensional analogue of the $1+1$ 
sigma model case (\ref{chiral},\ref{exp}) is the sigma model with coupling
\begin{eqnarray}
{\cal L}_{\rm int}&=&m\bar{\psi}\left(\pi_0+i\gamma_5 \vec{\pi}\cdot\vec{\tau}
\right)\psi \nonumber\\
&=& m\bar{\psi}\left(\frac{1}{2}(g+g^\dagger)+
\frac{1}{2}(g-g^\dagger)\gamma_5\right)\psi
\label{skyrme}
\end{eqnarray}
where $\vec{\tau}$ are $su(2)$ generators, the fields $\pi_0$ and $\vec{\pi}$ 
are constrained by $\pi_0^2+\vec{\pi}^2=1$, and the fields $g$ in the
second  line are defined by $g=\pi_0+i\vec{\pi}\cdot\vec{\tau}$. At zero
temperature, there is an induced topological charge density
\begin{equation}
J^0=\frac{1}{24\pi^2}\epsilon^{ijk}\, \tr\left(g^{-1}\partial_i g\,
g^{-1}\partial_j g\, g^{-1}\partial_k g\right)
\end{equation}
The corresponding zero temperature integrated charge is given by the winding 
number of the background field $g$ at zero temperature. We conjecture that 
at finite temperature this induced charge will acquire additional 
nontopological contributions similar to those found here for the $1+1$
sigma model case.

\vskip .1cm
{\bf Acknowledgements:}

This work has been supported in part (GD) by the U.S. Department of Energy
grant DE-FG02-92ER40716.00, and by PPARC. GD thanks Balliol College, Oxford,
for a Visiting Fellowship, and the Theoretical Physics Department at Oxford, 
and the CSSM at Adelaide for their hospitality and support.
\vskip 1cm



\begin{thebibliography}{99}
\bibitem{comment} This is similar to, but slightly different from, the 
point-monopole case considered in \cite{goldhaber} where the theta 
parameter is related to the boundary conditions used in the point limit.
\end{thebibliography}
\end{document}